# NUMERICAL COINCIDENCES AND 'TUNING' IN COSMOLOGY

MARTIN J. REES

**Abstract.** Fred Hoyle famously drew attention to the significance of apparent coincidences in the energy levels of the carbon and oxygen nucleus. This paper addresses the possible implications of other coincidences in cosmology.

## 1. Introduction

Hermann Bondi's classic book 'Cosmology' was, for many of us, an inspiring introduction to the science of the cosmos. In a chapter entitled 'Microphysics and Cosmology', Bondi lists the famous dimensionless constants, and mentions the well-known coincidence, first highlighted by Dirac, between the ratio of the electrical and gravitational forces within a hydrogen atom and the ratio of the Hubble radius to the size of an electron. He says: 'These coincidences are very striking, and few would deny their possible deep significance, but the precise nature of the connexion they indicate is not understood and is very mysterious.'

I am not sure to what extent Fred Hoyle was influenced in this matter by Bondi, but he certainly took this problem seriously too. He also, through his famous realisation of the $C^{12}$ resonance level's significance, made a celebrated addition to the list of cosmic coincidences. Moreover, in pondering their significance he was led to conjecture that the so called 'constants of nature' might not be truly universal. In 'Galaxies, Nuclei and Quasars' Hoyle writes that 'one must at least have a modicum of curiosity about the strange dimensionless numbers that appear in physics.' He goes on to outline two possible attitudes to them. One is that 'the dimensionless numbers are all entirely necessary to the logical consistency of physics'; the second possibility is that the numbers are not in the broadest sense universal, but that 'in other places their values would be different' Hoyle favoured this latter option because then 'the curious placing of the levels in $C^{12}$ and $O^{16}$ need no longer have the appearance of astonishing accidents. It could simply be that since creatures like ourselves depend on a balance between carbon and oxygen, we can exist only in the portions of the universe where these levels happen to be correctly placed.'

With these texts as my motivation, I'd like to summarise briefly how the issue looks today. I believe that Fred's conjecture is now even more attractive, though the 'portions of the universe' between which the variation occurs must now be interpreted as themselves vastly larger than the domain our telescopes can actually observe – perhaps even entire 'universes' within a multiverse.

But before dwelling further on these coincidences, it might be worth noting that the 'coincidence' that Dirac and Bondi discussed does not in itself now cause puzzlement. There is really just one very large number in physics: it is $e^2/Gm_p^2$ (or, equivalently, the reciprocal of the 'gravitational fine structure constant' $\alpha_G^{-1} = \left(\frac{\hbar c}{Gm_p^2}\right)$ which is larger by 137. The Chandrasekhar mass exceeds the proton mass by $\alpha_G^{-3/2}$. Stars are so large because gravity is so weak: Dicke (1961) also realised that they are also long-lived for the same reason. To present Dicke's estimate for stellar lifetimes in a slightly different way, we can define a characteristic time (cf Salpeter 1964) equal to

$$t_s = \frac{M_* c^2}{L_{\text{Ed}}} = \frac{2}{3}\left(\frac{e^2}{m_e c^3}\right)\left[\left(\frac{e^2}{Gm_p^2}\right)\left(\frac{m_p}{m_e}\right)\right]$$

This is the time it would take a body to radiate its rest mass energy if had the 'Eddington luminosity' where where radiation pressure balances gravity, and if electron scattering provided the main opacity. The lifetime of an actual star is obtained by multiplying $t_s$ by by various factors: the efficiency of nuclear energy ($\sim$ 0.007); the ratio of total pressure to radiation pressures ($> 1$); and the ratio of actual opacity to electron-scattering opacity ($> 1$). However, the key point (evident from the second way I have written the expression for $t_s$ above) is that stellar lifetimes are longer than the light travel time across the electron by the factor in square brackets which involves Dirac's large number. If we are observing the universe when its age is of order the age of a star (and there is such a time in a big bang model) then Dirac's 'coincidence' would naturally be satisfied.

## 2. Do the 'Special' Values of the Constants Need an Explanation?

If we ever established contact with intelligent aliens, how could we bridge the 'culture gap'? One common culture (in addition to mathematics) would be physics and astronomy. We and the aliens would all be made of atoms, and we'd all trace our origins back to the 'big bang' 13.7 billion years ago. We'd all share the potentialities of a (perhaps infinite) future. But our existence (and that of the aliens, if there are any) depends on our universe being rather special. Any universe hospitable to life – what we might call a *biophilic universe* – has to be 'adjusted' in a particular way. The prerequisites for any life of the kind we know about — long-lived stable stars, stable atoms such as carbon, oxygen and silicon, able to combine into complex molecules, etc — are sensitive to the physical laws and to the size, expansion rate and contents of the universe. Indeed, even for the most openminded science fiction writer, 'life' or 'intelligence' requires the emergence of some generic complex structures: it can't exist in a homogeneous universe, not in a universe containing only a few dozen particles. Many recipes would lead to stillborn universes with no atoms, no chemistry, and no planets; or to universes too short-lived or too empty to allow anything to evolve beyond sterile uniformity.

Consider, for example, the role of gravity. Stars and planets depend crucially on this force; however, we could not exist if gravity were much stronger than it actually is. A large, long-lived and stable universe depends quite essentially on $\alpha_G^{-1}$ being exceedingly large. Gravity also amplifies 'linear' density contrasts in an expanding universe; it then provides a negative specific heat so that dissipative bound systems heat up further as they radiate. There's no thermodynamic paradox in evolving from an almost structureless fireball to the present cosmos, with huge temperature differences between the 3 degrees of the night sky, and the blazing surfaces of stars. So gravity is crucial, but the weaker it is, the grander and more prolonged are its consequences.

Newton's constant G need not be fine-tuned – merely exceedingly weak so that $\alpha_G^{-1}$ is indeed very large. However, the natural world is much more sensitive to the balance between other basic forces. If nuclear forces were slightly stronger than they actually are relative to electric forces two protons could stick together so readily that ordinary hydrogen would not exist, and stars would evolve quite differently. Some of the details are still more sensitive, as Hoyle emphasised.

Even a universe as large as ours could be very boring: it could contain just black holes, or inert dark matter, and no atoms at all. Even if it had the same ingredients as ours, it could be expanding so fast that no stars or galaxies had time to form; or it could be so turbulent that all the material formed vast black holes rather than stars or galaxies. – an inclement environment for life. And our universe is also special in having three spatial dimensions. A four dimensional world would be unstable; in two dimensions, nothing complex could exist.

The distinctive and special-seeming recipe characterising our universe seems to me a fundamental mystery that should not be brushed aside merely as a brute fact. Rather than re-addressing the classic 'fine tuning' examples, I shall focus on the parameters of the big bang – the expansion rate, the curvature, the fluctuations, and the material content. Some of these parameters (perhaps even all) may be explicable in terms of a unified theory: or they may be somehow derivable from the microphysical constants. But, irrespective of how that may turn out, it is interesting to explore the extent to which the properties of a universe – envisaged here as the aftermath of a single big bang – are sensitive to the cosmological parameters.

### 3. The Cosmological Numbers

Traditionally, cosmology was the quest for a few numbers. The first were H, and q. Since 1965 we've had another : the baryon/photon ratio $n_b/n_\gamma$. This is believed to result from a small favouritism for matter over antimatter in the early universe – something that was addressed in the context of 'grand unified theories' in the 1970s. (Indeed, baryon non-conservation seems a prerequisite for any plausible in-flationary model. Our entire observable universe, containing at least $10^{79}$ baryons, could not have inflated from something microscopic if baryon number were strictly conserved).

In the 1980s non-baryonic matter became almost a natural expectation, and $\Omega_b/\Omega_{DM}$ is another fundamental number.

We now have the revival of the cosmological constant lambda (or some kind of 'dark energy', with negative associated pressure, which is generically equivalent to lambda).

Another specially important dimensionless number tells us how smooth the universe is. It's measured by

- The Sachs-Wolfe fluctuations in the microwave background
- the gravitational binding energy of clusters as a fraction of their rest mass
- or by the square of the typical length scale of mass- clustering as a fraction of the Hubble radius.

It's of course oversimplified to represent this by a single number, but insofar as one can, its value (let's call it Q) is pinned down to be $10^{-5}$. (Detailed modelling of the fluctuations introduces further numbers: the ratio of scalar and tensor amplitudes, and quantities such as the 'tilt', which measure the deviation from a pure scale-independent Harrison-Zeldovich spectrum.)

### 4. Anthropic Requirements for a Universe

We can make a list of what would be required for a big bang to yield an 'anthropically allowed' universe – a universe where complexity, whether humanoid or more like a black cloud, could unfold. The list would include the following:

- Some inhomogeneities (i.e. a non-zero Q): clearly there is no potential for complexity if everything remains in a uniform ultra-dilute medium
- Some baryons: complexity would be precluded in a universe solely made of dark matter, with only gravitational interactions.
- At least one star (probably, though perhaps superfluous for black-cloud-style complexity)
- Some second-generation stars: only later-generation stars would be able to have orbiting planets, unless heavy elements were primordial.

It is interesting to engage in 'counterfactual history' and ask what constraints these various requirements would impose on hypothetical universes with different characteristics – in particular, with different values of:

- The fluctuation amplitude Q (which is $10^{-5}$ in our actual universe)
- The cosmological constant
- The baryon/photon ratio (about $10^{-9}$ in our universe)
- The baryon/dark matter ratio $\Omega_b/\Omega_{DM}$ (about 0.2 in our universe).

## 4.1. THE FLUCTUATION AMPLITUDE

First, we might explore what a universe would be like which was initially smoother (Q smaller) or rougher (Q larger) than ours.

Were Q of order $10^{-6}$, there would be no clusters of galaxies; moreover, the only galaxies would be small and anaemic. They would form much later than galaxies did in our actual universe. Because they are loosely bound, processed material would be expelled from shallow potential wells; there may therefore be no second-generation stars, and so no planetary systems. If Q were even smaller than $10^{-6}$, there would be no star formation at all: very small structures of dark matter would turn around late, and their constituent gas would be too dilute to undergo the radiative cooling that is a prerequisite for star formation. (In a lambda-dominated universe, isolated clumps could survive for an infinite time without merging into a larger scale in the hierarchy. So eventually, for any $Q > 10^{-8}$, a 'star' could form – but by that time it might be the only bound object within the horizon).

Hypothetical astronomers in a universe with $Q = 10^{-4}$ might find their cosmic environment more varied and interesting than ours. Galaxies and clusters would span a wider range of masses. The biggest clusters would be 30 times more massive than any in our actual universe. There could be individual 'galaxies' – perhaps even disc galaxies – with masses up to that of the Coma cluster and internal velocity dispersions up to 2000 km/sec. These would have condensed when the universe was only $3.10^8$ years old, and when Compton cooling on the microwave background was still effective

However a universe where Q were larger still – more than (say) $10^{-3}$ – would be a violent and inhospitable place. Huge gravitationally-bound systems would collapse, trapping their radiation and unable to fragment, soon after the epoch of recombination. (Collapse at, say, $10^7$ years would lead to sufficient partial ionization (via strong shocks) to recouple the baryons and the primordial radiation.) Such structures, containing the bulk of the material, would turn into vast black holes. It is unlikely that galaxies of any kind would exist; nor is it obvious that much baryonic material would ever go into stars: even if so, they would be in very compact highly bound systems).

(Note that, irrespective of these anthropic constraints on its value, Q has to be substantially less than one in order to make cosmology a tractable subject, separate from astrophysics, This is because the ratio of the largest structures to the Hubble radius is of order $Q^{1/2}$. Numbers like $\Omega$ and H *are only well-defined* insofar as the universe possesses 'broad brush' homogeneity – so that our observational horizon encompasses many independent patches each big enough to be a fair sample. This wouldn't be so, and the simple Friedmann models wouldn't be useful approximations, if Q *weren't* much less than unity.)

According to most theories of the ultra-early universe, Q is imprinted by quantum effects: microscopic fluctuations, after exponential expansion, give rise to the large-scale irregularities observed in the microwave background sky, and which are the 'seeds' for galaxies and clusters. But as yet no theories pin down Q's value.

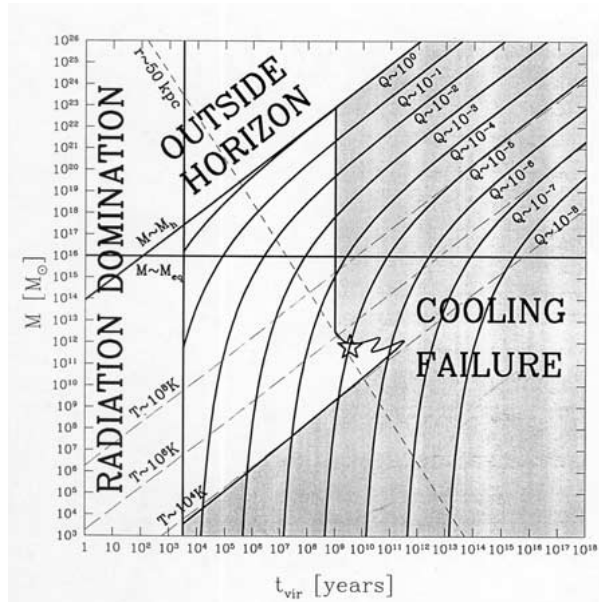

*Figure 1.* The domains in the which bound structures can form, for different values of Q (from Tegmark and Rees, 1998).

### 4.2. Λ OR DARK ENERGY

Theorists are even further from understanding $\Lambda$. Indeed, the naive guess is that $\Lambda$ should be least 120 powers of 10 larger than it could be in our actual universe – unless there were some cancellation mechanism. (Indeed, inflation models postulate an effective vacuum density that was indeed as high as this for a brief initial interval.)

The interest has of course been hugely boosted recently, through the convergence of several lines of evidence on a model where the universe is close to being 'flat', but with 4 percent in baryons, about 25 percent in dark matter, and the remaining (dominant) component in $\Lambda$ or some time-dependent 'dark energy'. (Incidentally, the full resurrection of $\Lambda$ would be a great 'coup' for de Sitter. His model, dating for the 1920s, not only describes inflation, but would then also describes future aeons of our cosmos with increasing accuracy. Only for the 50-odd decades of logarithmic time between the end of inflation and the present would it need modification!). For a universe with the actual observed values of $Q$, it is readily shown that a value of $\Lambda$ more than 5-10 times higher than the apparent 'dark energy' density would have the 'anti-anthropic' consequence of precluding galaxy formation. This happens because the cosmic repulsion would then be so fierce that it would take over before any galaxies had a chance to form via gravitational instability.

### 4.3. $\Omega_b$ AND $\Omega_{DM}$: THE BARYON/DARK MATTER DENSITY

Baryons are anthropically essential; there are firm lower limits on their requisite abundances, but they need not be the dominant constituent. (Indeed they are far from dominant in our actual universe). Lower $n_b/n_\gamma$ and lower $\Omega_{DM}$ reduce the 'efficient cooling' domain in the Rees/Tegmark (1998) curves. reproduced in Figure 1.

If the photons outnumbered the baryons and the dark matter particles by a still larger factor than in our actual universe, then the universe would remain radiation-dominated for so long that the gravitational growth of fluctuations would be inhibited (Rees, 1980).

On the other hand, a higher value of $n_b/n_\gamma$ $(1 + \Omega_{DM}/\Omega_b)$ reduces $t_{eq}$ and allows gravitational clustering to start earlier. This reduces the minimum Q required for emergence of non-linear structures (cf. Aguirre, 2001)

(Note also that the mechanism that gives rise to baryon favouritism may be linked to the strong interactions, and therefore correlate with key numbers in nuclear physics.)

### 4.4. DELINEATING THE ANTHROPICALLY-ALLOWED DOMAIN

In the above, I have envisaged changing just one parameter, leaving the others with their actual values. But of course there may be correlations between them. For example, suppose that there were big bangs with a whole range of Q-values. Structures form earlier (when the matter density is higher) in universes with larger Q, so obviously a higher Q is anthropically-compatible with a higher $\Lambda$.

If we consider a two-dimensional situation where Q and $\Lambda$ vary, then we find that there is an anthropically allowed area. There are (rather vaguely defined) upper and lower limits to Q (as already discussed) but within the range, there is an upper limit to Q (see Figure 2).

We can carry out the exercise, in as many dimensions as we wish, of delineating the anthropically-allowed domain in parameter space. (even though to quantify this is more difficult). To delineate the allowed domains is procedurally uncontroversial, but what about the motivation? It obviously depends on believing that the laws of nature could have been otherwise – unless there is some scientific validity in imagining 'counterfactual universes' this exercise seems vacuous.

## 5. Is it 'Scientific' to enquire about other Universes?

If our existence – or, indeed, the existence of any 'interesting' universe – depends on a seemingly special cosmic recipe, how should we react? There seem two lines to take: we can dismiss it as happenstance, or we can conjecture that our universe is a specially favoured domain in a still vaster multiverse.

(a) *Happenstance (or coincidence)*

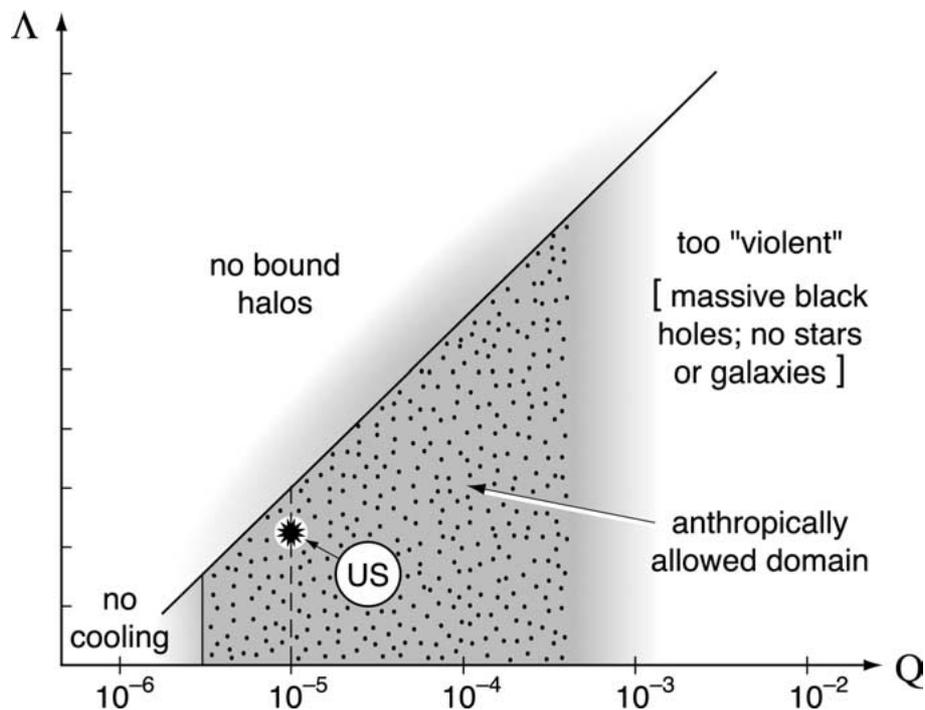

*Figure 2.* This shows in a two-dimensional parameter space Λ and Q. The upper and lower limits to Q are discussed by Tegmark and Rees (1998). The upper limit to Λ stems from the requirement that galactic-mass bound systems should form before the universe enters its accelerating de Sitter phase. Our universe (obviously) lies in the anthropically-allowed domain. But we cannot say whether it is at a typical location within that domain without a specific model for the probability distributions of Q and Λ in the ensemble.

Maybe a fundamental set of equations, which some day will be written on T-shirts, fixes all key properties of our universe uniquely. It would then be an unassailable fact that these equations permitted the immensely complex evolution that led to our emergence.

But I think there would still be something to wonder about. It's not guaranteed that simple equations permit complex consequences To take an analogy from mathematics, consider the Mandelbrot set. This pattern is encoded by a short algorithm, but has infinitely deep structure: tiny parts of it reveal novel intricacies however much they are magnified. In contrast, you can readily write down other algorithms, superficially similar, that yield very dull patterns. Why should the fundamental equations encode something with such potential complexity, rather than the boring or sterile universe that many recipes would lead to?

One hard-headed response is that we couldn't exist if the laws had boring consequences. We manifestly are here, so there's nothing to be surprised about. I think we would need to know why the unique recipe for the physical world should permit

consequences as interesting as those we see around us ( and which, as a byproduct, allowed us to exist)

(b) *A special universe drawn from an ensemble, or multiverse*

But there is another perspective – a highly speculative one, however. There may be many 'universes' of which ours is just one. In the others, some laws and physical constants would be different. But our universe wouldn't be just a random one. It would belong to the unusual subset that offered a habitat conducive to the emergence of complexity and consciousness. If our universe is selected from a multiverse, its seemingly designed or fine tuned features wouldn't be surprising. This might seem arcane stuff, disjoint from 'traditional' cosmology – or even from serious science. But my prejudice is to be openminded about ensembles of universe and suchlike, and even to suspect that we may not be able to account for some features of our own universe without invoking them.

First, a semantic digression: the word 'universe' traditionally denotes 'everything there is'. Therefore if we are to consider other domains of space time (originating in other big bangs) we should really define the whole ensemble as 'the universe', and introduce a new word – 'metagalaxy' for instance – to denote what observational cosmologists traditionally study. However, so long as this whole idea remains speculative, it is probably best to continue to denote what cosmologists observe as 'the universe', and to introduce a new term, 'multiverse', for the whole hypothetical ensemble.

Some might regard other universes — regions of space and time that we cannot observe (perhaps even in principle and not just in practice) – as being in the province of metaphysics rather than physics. Science is an experimental or observational enterprise, and it's natural to be troubled by invocations of something unobservable. But I think 'other universes' (in this sense) already lie within the proper purview of science. It is not absurd or meaningless to ask 'Do unobservable universes exist?', even though no quick answer is likely to be forthcoming. The question plainly can't be settled by *direct* observation, but relevant evidence *can* be sought, which could lead to an answer.

There is actually a blurred transition between the readily observable and the absolutely unobservable, with a very broad grey area in between (see Figure 3). To illustrate this, one can envisage a succession of horizons, each taking us further than the last from our direct experience:

(i) *Limit of present-day telescopes*

There is a limit to how far out into space our present-day instruments can probe. Obviously there is nothing fundamental about this limit: it is constrained by current technology. Many more galaxies will undoubtedly be revealed in the coming decades by bigger telescopes now being planned. We would obviously not demote such galaxies from the realm of proper scientific discourse simply because they haven't been seen yet.

(ii) *Limit in principle at present era*

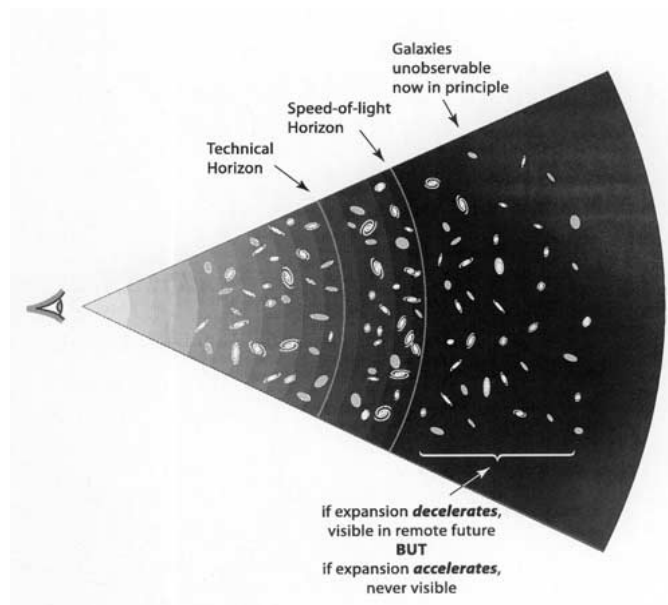

*Figure 3.* Extending horizons beyond the directly-observable.

Even if there were absolutely no technical limits to the power of telescopes, our observations are still bounded by the particle horizon, which demarcates the spherical shell around us at which the redshift would be infinite.

If our universe were decelerating, then the horizon of our remote descendants would encompass extra galaxies that are beyond our horizon today. It is, to be sure, a practical impediment if we have to await a cosmic change taking billions of years, rather than just a few decades (maybe) of technical advance, before a prediction about a particular distant galaxy can be put to the test. But does that introduce a difference of principle? Surely the longer waiting-time is a merely quantitative difference, not one that changes the epistemological status of these faraway galaxies?

(iii) *Never-observable galaxies from 'our' Big Bang*

But what about galaxies that we can *never* see, however long we wait? It's now believed that we inhabit an accelerating universe. As in a decelerating universe, there would be galaxies so far away that no signals from them have yet reached us; but if the cosmic expansion is accelerating, we are now receding from these remote galaxies at an ever-increasing rate, so if their light hasn't yet reached us, it never will. Such galaxies aren't merely *unobservable in principle now* – they will be beyond our horizon *forever*. But if a galaxy is now unobservable, it hardly seems to matter whether it remains unobservable for ever, or whether it would come into view if we waited a trillion years. (And I have argued, under (ii) above, that the latter category should certainly count as 'real'.)

(iv) *Galaxies in disjoint universes*

The never-observable galaxies in (iii) would have emerged from the same Big Bang as we did. But suppose that, instead of causally-disjoint regions emerging from a single Big Bang (via an episode of inflation) we imagine separate Big Bangs. Are space-times completely disjoint from ours any less real than regions that never come within our horizon in what we'd traditionally call our own universe? Surely not – so these other universes too should count as real parts of our cosmos, too

Whether other universes exist or not is a scientific question. Those who are prejudiced against the concept should regard the above step-by-step argument as an exercise in 'aversion therapy'. From a reluctance to deny that galaxies with redshift 10 are proper objects of scientific enquiry, you are led towards taking seriously quite separate space-times, perhaps governed by quite different 'laws'.

Linde, Vilenkin and others have performed computer simulations depicting an 'eternal' inflationary phase where many universes sprout from separate big bangs into disjoint regions of spacetimes – each such region itself vastly larger than our observational horizon. Guth, Harrison and Smolin have, from different viewpoints, suggested that a new universe could sprout inside a black hole, expanding into a new domain of space and time inaccessible to us. And Randall and Sundrum suggest that other universes could exist, separated from us in an extra spatial dimension; these disjoint universes may interact gravitationally, or they may have no effect whatsoever on each other.

None of these scenarios has been simply dreamed up out of the air: each has a serious, albeit speculative, theoretical motivation. However, one of them, at most, can be correct. Quite possibly none is: there are alternative theories that would lead just to one universe. Firming up any of these ideas will require a theory that consistently describes the extreme physics of ultra-high densities, how structures on extra dimensions are configured, etc. Perhaps, in the 21st-century theory, physicists will develop a theory that yields insight into (for instance) why there are three kinds of neutrinos, and the nature of the nuclear and electric forces. Such a theory would thereby acquire credibility. If the same theory, applied to the very beginning of our universe, were to predict many big bangs, then we would have as much reason to believe in separate universes as we now have for believing inferences from primordial nucleosynthesis about the first few minutes of cosmic history.

## 6. Universal Laws, or Mere Bylaws?

Some theorists, Frank Wilczek for instance, regard 'are the laws of physics unique?' as a key scientific challenge for the new century. The answer determines how much variety the other universes – if they exist – might display. If there were something uniquely self-consistent about the actual recipe for our universe, then the aftermath of any big bang would be a re-run of our own universe. But a far more interesting possibility (which is certainly tenable in our present state of ignorance of the underlying laws) is that *the underlying laws governing the entire multiverse may allow variety among the universes.* Some of what we call 'laws of nature' may in this grander perspective be *local bylaws*, consistent with some overarching theory governing the ensemble, but not uniquely fixed by that theory. Many things in our cosmic environment – for instance, the exact layout of the planets and asteroids in our Solar System – are accidents of history. Likewise, the recipe for an entire universe may be arbitrary.

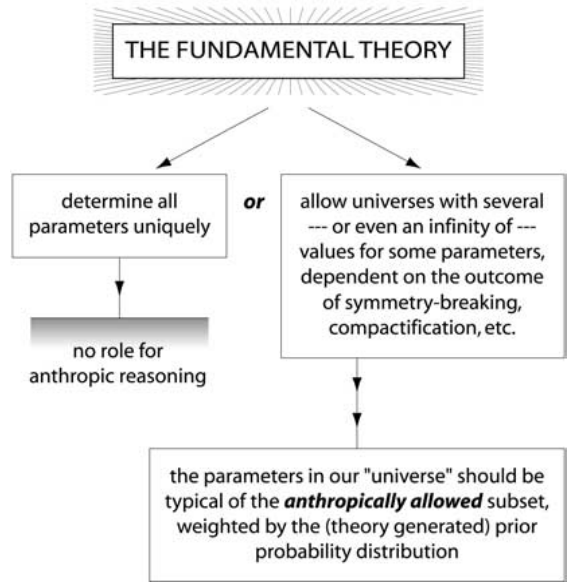

*Figure 4.* 'Decision tree'. Progress in 21-st century physics should allow us to decide whether anthropic explanations are irrelevant or, on the other hand, the best we can ever hope for.

More specifically, some aspects may be arbitrary and others not. As an analogy (which I owe to Paul Davies) consider the form of snowflakes. Their ubiquitous six-fold symmetry is a direct consequence of the properties and shape of water molecules. But snowflakes display an immense variety of patterns because each is moulded by its micro-environments: how each flake grows is sensitive to the fortuitous temperature and humidity changes during its growth.

If physicists achieved a fundamental theory, it would tell us which aspects of nature were direct consequences of the bedrock theory (just as the symmetrical template of snowflakes is due to the basic structure of a water molecule) and which are (like the distinctive pattern of a particular snowflake) the outcome of accidents.

The cosmological numbers in our universe, and perhaps some of the so-called constants of laboratory physics as well, could be 'environmental accidents', rather than uniquely fixed throughout the multiverse by some final theory. Some seemingly 'fine tuned' features of our universe could then only be explained by 'anthropic' arguments [see Figure 4]. Although this style of explanation raises hackles among some physicists it is analogous to what any observer or experimenter does when they allow for selection effects in their measurements: if there are many universes, most of which are not habitable, we should not be surprised to find ourselves in one of the habitable ones!

The entire history of our universe could just be an episode of the infinite multiverse; what we call the laws of nature (or some of them) may be just parochial bylaws in our cosmic patch. Such speculations dramatically enlarge our concept of reality. Putting them on a firm footing must await a successful fundamental theory that tells us whether there could have been many 'big bangs' rather than just one, and (if so) how much variety they might display. Until this fundamental issue is settled one way or the other, we won't know whether anthropic arguments are irrelevant or unavoidable.

### 7. Testing Multiverse Theories Here and Now

We may one day have a convincing theory that accounts for the very beginning of our universe, tells us whether a multiverse exists, and (if so) whether some so called laws of nature are just parochial by-laws in our cosmic patch. But while we're waiting for that theory – and it could be a long wait – we can check whether anthropic selection offers a tenable explanation for the apparent fine tuning. Such a hypothesis could even be refuted: this would happen if our universe turned out to be *even more specially* tuned than our presence requires.

We could apply this style of reasoning to the important numbers of physics (for instance, $\Lambda$) to test whether our universe is typical of the subset that that could harbour complex life. Most physicists would consider the 'natural' value of $\Lambda$ to be large, because it is a consequence of a very complicated microstructure of space. Perhaps there is only a rare subset of universes where $\Lambda$ is below the threshold that allows galaxies and stars to form. $\Lambda$ in *our* universe obviously had to be below that threshold, But if our universe were drawn from an ensemble in which $\Lambda$ was equally likely to take any value, we wouldn't expect it to be *too far below it*.

Current evidence suggests that if $\Lambda$ constituted the 'dark energy', its actual value is 5-10 times below that threshold. That would put our universe between the 10th or 20th percentile of universes in which galaxies could form. In other words, our universe isn't significantly more special, with respect to $\Lambda$, than our emergence demanded. But suppose that (contrary to current indications) observations showed that $\Lambda$ made no discernible contribution to the expansion rate, and was *thousands of times* below the threshold, not just 5–10 times. This 'overkill precision' would raise doubts about the hypothesis that $\Lambda$ was equally likely to have any value, and suggest that it was zero for some fundamental reason (or that it had a discrete set of possible values, and all the others were well about the threshold).

I've taken $\Lambda$ just as an example. We could analyse other important numbers of physics in the same way, to test whether our universe is typical of the habitable subset that that could harbour complex life. The methodology requires us to decide what values are compatible with our emergence. It also requires a specific theory that gives the probability of any particular value.

With this information, one can then ask if our actual universe is 'typical' of the subset in which we could have emerged. If it is an atypical member even of this subset (not merely of the entire multiverse) then our hypothesis would be disproved. Other parameters could be analysed similarly – testing in a multi-parameter space whether our universe is a typical member within the anthropically allowed domain.

As a two-dimensional example, consider the joint constraints on $\Lambda$ and Q in Figure 2. We cannot decide whether our universe is typical without a theory that tells us what 'measure' to put on each part of the 2-dimensional parameter space. If high-Q universes were more probable, and the probability density of $\Lambda$ were uniform, then we should be surprised not to find ourselves in a universe with higher $\Lambda$ and higher Q.

These examples show that some claims about other universes may be refutable, as any good hypothesis in science should be. We cannot confidently assert that there were many big bangs – we just don't know enough about the ultra-early phases of our own universe. Nor do we know whether the underlying laws are 'permissive': settling this issue is a challenge to 21st century physicists. But if they are, then so-called anthropic explanations would become legitimate – indeed they'd be *the only type of explanation we'll ever have* for some important features of our universe.

Models with low omega, non-zero lambda two kinds of dark matter, and the rest may *seem* ugly. Some theorists are upset by these developments, because it frustrates their craving for maximal simplicity. I think we can learn a lesson from cosmological debates in the 17th century. Galileo and Kepler were upset that planets moved in elliptical orbits, not in perfect circles. Newton later showed, however, that all elliptical orbits could be understood by a single unified theory of gravity. Likewise our universe may be just one of an ensemble of all possible universes, constrained only by the requirement that it allows our emergence. But to regard this outcome as ugly may be as myopic as Kepler's infatuation with circles: Newton was perhaps the greatest scientific intellect of the second millennium. Perhaps his third-millennium counterpart will uncover a mathematical system that governs the entire multiverse.